\documentclass[twocolumn]{aastex631}
\usepackage[normalem]{ulem}  
\usepackage[T1]{fontenc}
\usepackage{ae,aecompl}
\usepackage{graphicx}
\usepackage{amsmath}
\usepackage{amssymb}
\usepackage{supertabular}
\usepackage{hyperref}
\usepackage{xcolor}
\usepackage{academicons}
\usepackage{multirow}

\newcommand{\Msun}{\,{M}_{\odot}}
\newcommand{\km}{\hbox{$\,{\rm km}$}}
\newcommand{\D}{{\rm d}}

\def\Hz{\;\text{Hz}}

\begin{document}
\title{Resolving phase transition properties of dense matter through tidal-excited g-mode from inspiralling neutron stars}

\author[0000-0003-1197-3329]{Zhiqiang Miao}
\affiliation{Department of Astronomy, Xiamen University, Xiamen, Fujian 361005, China}

\author[0000-0002-9624-3749]{Enping Zhou}
\affiliation{Department of Astronomy, School of Physics, Huazhong University of Science and Technology, Wuhan, 430074, China}

\author[0000-0001-9849-3656]{Ang Li}
\affiliation{Department of Astronomy, Xiamen University, Xiamen, Fujian 361005, China}

\correspondingauthor{Ang Li}
\email{liang@xmu.edu.cn}

\begin{abstract} 
The investigation of the phase state of dense matter is hindered by complications of first-principle nonperturbative quantum chromodynamics. By performing the first consistent general-relativistic calculations of tidal-excited g-mode of neutron stars with a first-order strong interaction phase transition in the high-density core, we demonstrate that gravitational wave signal during binary neutron star inspiral probes their innermost hadron-quark transition and provides potent constraints from present and future gravitational-wave detectors. 
\end{abstract}


\section{Introduction}
Despite decades of exploration, the phase state of cold, dense matter is still a fundamental problem that remains unresolved. This problem pertains to low-energy nonperturbative quantum chromodynamics (QCD)~\citep{2010PhRvD..81j5021K} and is one of the key scientific goals in current and anticipated relativistic heavy-ion collision experiments~\citep{2022Natur.606..276H}. It is widely believed that this type of matter exists in the core of neutron stars (NSs)~\citep{1999paln.conf.....W,2000ARNPS..50..481H,2001ApJ...550..426L}. The equation of state (EOS) of dense stellar matter has drawn much attention in the era of gravitational wave (GW) astronomy. This is because the observations of GW170817 led to the first measurement of NS tidal deformability~\citep{2017PhRvL.119p1101A}, which offered valuable insight into the phase state of dense matter~\citep{2018PhRvL.121p1101A}. 

The tidal deformability accounts for the quadrupole deformations of inspiralling NSs, which are induced by the tidal force from each other, when the quasi-equilibrium approximation can be utilized (i.e., the tidal field changes slowly). While in the late stage of the inspiral phase, once the binary's orbital frequency approaches the frequency of a particular normal mode of NSs, a resonance can occur. This resonance will transfer the orbital energy to the stellar oscillation via a mechanism referred to as the dynamical tide. This event will accelerate the orbital decay and result in a phase shift in the GW waveform, which will provide insights into the NS inner structure with the help of GW detection from inspiralling NSs.

Previous studies have focused on the interfacial (i) modes, which are excited at the interface between the solid crust and fluid core~\citep{2012PhRvL.108a1102T,2020PhRvL.125t1102P,2023PhRvD.107h3023Z}. These modes typically have frequencies of $\sim 100\Hz$ and contribute $\sim\mathcal{O}(0.1)$ phase shifts during resonance, making them promising candidates for detection using Advanced LIGO (aLIGO)~\citep{2015CQGra..32g4001L} and 3rd generation GW detectors such as Einstein Telescope (ET)~\citep{2020JCAP...03..050M} and Cosmic Explorer (CE)~\citep{2019BAAS...51g..35R}. Attempts have also been made to infer NS EOS through tide-induced f-modes and g-modes, which arise from the composition gradients within the NS's interior. Studies~\citep[see e.g.,][]{1999LRR.....2....2K,1999MNRAS.308..153H,1994ApJ...426..688R,1994MNRAS.270..611L,2017MNRAS.464.2622Y,2017PhRvD..96h3005X} have shown that f-modes are challenging to excite due to their high resonant frequencies, whereas g-modes demonstrate negligible phase shifts because their couplings to tidal fields are meager. 

In this work, we investigate a unique type of g-modes, known as discontinuity g-modes, which are expected to occur in NSs that possibly have a strong phase transition in their high-density inner cores~\citep{1987MNRAS.227..265F}\footnote{We study in the context of slow phase transition, namely the timescale of hadron-quark conversion is much longer than the typical oscillation timescale. An alternative fast conversion would result in a g-mode with zero frequency~\citep{2020PhRvD.101l3029T} and the present analysis will not apply.}. 
Our study showcases how discontinuity g-modes' dynamical tide can be leveraged to probe the dense matter's phase transition properties using the data obtained from GW170817, along with possible observations from the next-generation GW detectors.

\section{Formalism and methods}
\subsection{Relativistic g-modes and tidal coupling}
The existence of a quark core inside NSs (also called hybrid stars) has been discussed with Newtonian formulations of perturbations and tidal couplings, focusing on the i-mode at the interface of a crystalline quark core and a fluid hadronic envelope~\citep{2021PhRvD.103f3015L}.
With the solid quark core condition relaxed, we here employ a full general relativistic (GR) formalism as follows: 1) Constructing the background spherical hybrid stars in hydrostatic equilibrium using Einstein's field equations, 
i.e., solving the Tolman-Oppenheimer-Volkoff (TOV) equations, and 2) Computing their discontinuity g-modes by solving the linear pulsation equations in general relativity~\citep{1983ApJS...53...73L,1985ApJ...292...12D}. 
Our study focuses solely on the $l=2$ g-modes, as they have the strongest coupling to the tidal gravitational field. 

Our approach prioritizes maintaining consistency in how we treat the oscillation mode and the tidal overlap integral $Q$ within the framework of GR (see Appendix~\ref{appendix: tidal overlap} for a derivation of this form):
\begin{equation}\label{eq:rel Q}
    Q=\frac{1}{MR^2}\int\sqrt{-g}e^{-2\Phi}d^3x(p+\varepsilon)\xi^{*\mu}\nabla_\mu(r^2Y_{2m})\ .
\end{equation}
Here, $g$ represents the determinant of the metric of the background spacetime and $e^{2\Phi}=g_{tt}$ is the $(t,t)$ component of the metric. $Y_{2m}$ is the $l=2$ spherical harmonic. $\xi^\mu =(0,re^{-\lambda}WY_{2m},V\partial_\theta Y_{2m},V\sin^{-2}\theta\partial_\phi Y_{2m})e^{i\omega t}$ is the Lagrangian displacement of the eigenmode with the normalization $\langle \xi^\mu|\xi^\mu\rangle=MR^2$, where $e^{2\lambda}=g_{rr}$ is the $(r,r)$ component of the metric and $\langle \xi^\mu|\psi^\mu\rangle =\int\sqrt{-g}e^{-2\Phi}d^3x(p+\varepsilon)\xi^{*\mu}\psi_\mu$ is the relativistic inner product. $f$ is the frequency of the mode and $\omega=2\pi f$ is the angular frequency.
The present self-consistent approach effectively mitigates the issue of ``f-mode contamination''~\citep{1994ApJ...432..296R} in the calculation of $Q$ encountered in the hybrid (GR background + Newtonian perturbation) approach commonly used previously.
We provide the full description of this relativistic formula in Appendix~\ref{appendix: tidal overlap}, where we've also validated that our relativistic $Q$ satisfies both orthogonality and the sum rule.

We determine the mass-radius ($M$-$R$) relationship of NSs through the solution of the TOV equations, using the EOS - i.e., pressure ($p$) as a function of energy density ($\varepsilon$) - as our basic input.
To describe the fiducial soft and stiff hadronic matter (HM), we employ the QMF~\citep{2018ApJ...862...98Z} and NL3$\omega\rho$~\citep{2001PhRvL..86.5647H} models, ensuring their consistency with results from laboratory nuclear experiments.
For modeling the sharp first-order hadron-quark phase transition and representing quark matter within the high-density cores of NSs, we utilize the well-established generic constant-speed-of-sound (CSS) parameterization~\citep{2013PhRvD..88h3013A}.
Previous studies have thoroughly discussed the weak density dependence of the speed of sound in quark matter~\citep[see e.g.,][]{2013PhRvD..88h3013A,2015PhRvD..92h3002A}. 
This approach enables us to establish the EOS from the start of the phase transition up to the maximum central density of the star by using three phase-transition parameters: the transition pressure $p_{\rm t}$ (or the transition density $\varepsilon_{\rm t} \equiv \varepsilon_{\rm HM} (p_{\rm t})$), the transition strength $\Delta\varepsilon$ and the speed of sound in quark matter $c_s$. 
The transition density and transition strength affect the tidal overlap integral and its typical value remains at $\sim\mathcal{O}(0.01)$.

It is worth noting that since the discontinuity g-modes are a result of the discontinuity in the density profile of the star, we expect the frequency of these modes to depend on both the transition strength and the transition density. Based on our calculations, we observe that the g-mode frequency typically falls within the range of $100\Hz$ to $1500\Hz$, which is strongly dependent on the transition strength. The transition strength varies between $\sim5\,{\rm MeV/fm^3}$ and $\sim300\,{\rm MeV/fm^3}$.
Our study also indicates that the speed of sound ($c_s$) has a subdominant effect on mode frequency. Specifically, the difference in mode frequency between the $c_s=1/\sqrt3$ case (corresponding to the conformal limit) and the $c_s=1$ case (corresponding to the casual limit) is found to be less than $\sim10\%$. Thus, for the purpose of our analysis, we solely focus on extremely stiff quark matter with a speed of sound $c_s=1$.

\begin{figure*}
  \centering
    \centering
    \includegraphics[width=0.49\linewidth]{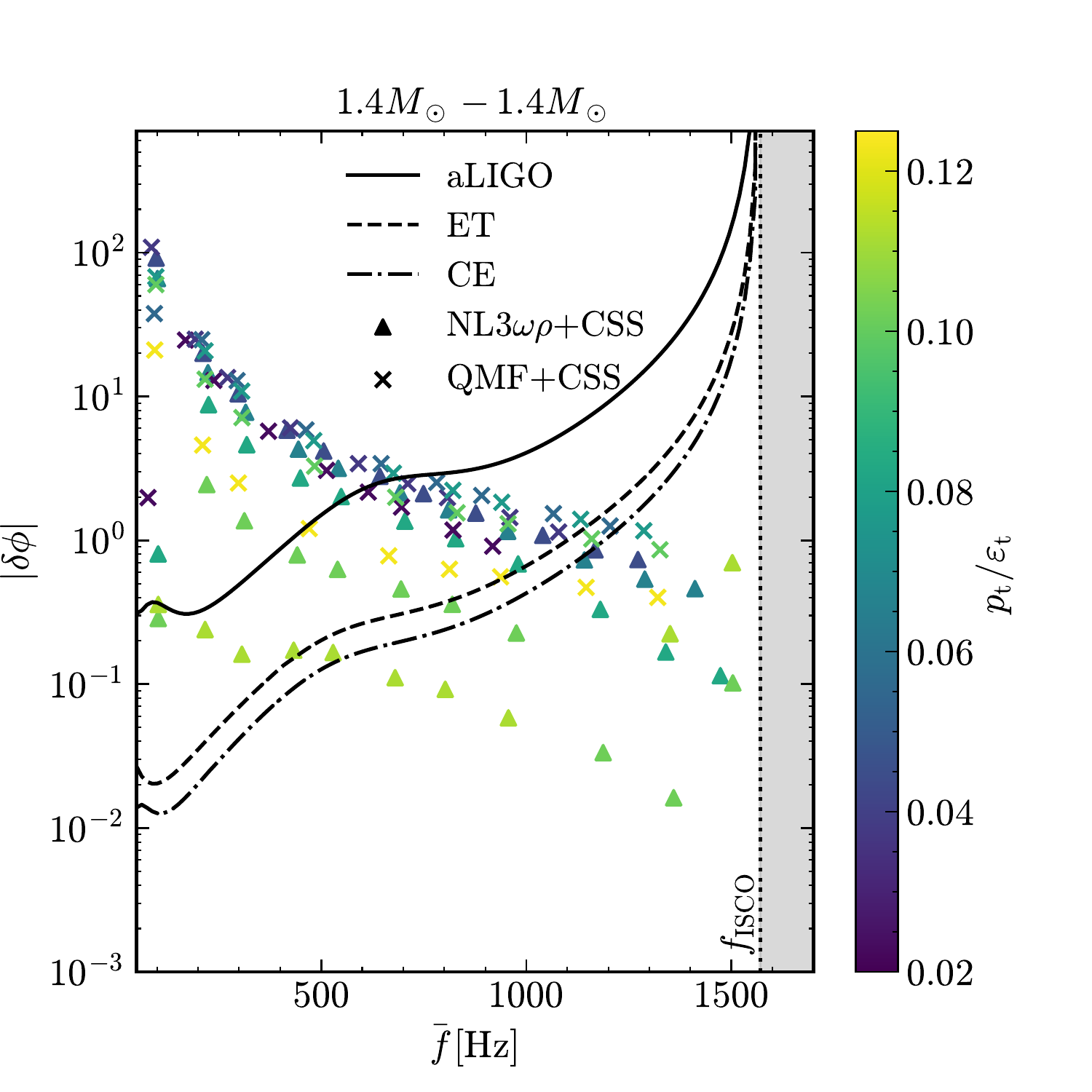}
    \centering
    \includegraphics[width=0.49\linewidth]{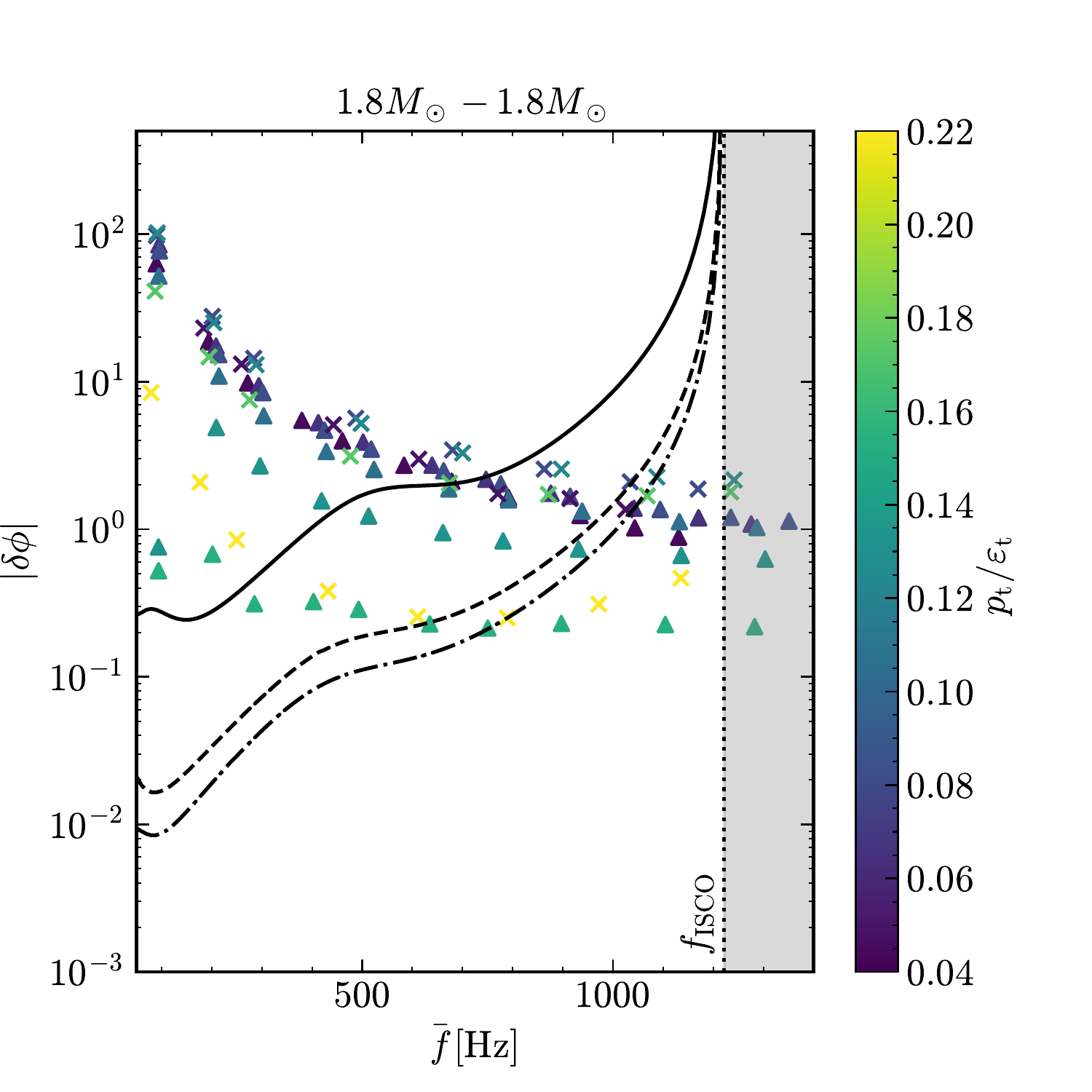}
    \caption{Total phase shift $|\delta\bar\phi|$ and resonant frequency $\bar f$ for a $1.4\Msun-1.4\Msun$ system (left panel) and a $1.8\Msun-1.8\Msun$ system (right panel), depending on the transition densities (represented by $p_{\rm t}/\varepsilon_{\rm t}$).
    $f_{\rm ISCO}\simeq 4400\Hz/(M+M^\prime)$ denotes the GW frequency corresponding to the innermost stable circular orbit, at which we assume the signal of inspiral phase shut off~\citep{1994PhRvD..49.2658C}.
     Also plotted are the detectability thresholds of aLIGO, ET and CE, which are calculated from their designed sensitivities by assuming the system is located at $D_L=100\,{\rm Mpc}$.
    } 
\label{fig:detectability}
\end{figure*}

\begin{figure}
  \centering
    \includegraphics[width=0.49\textwidth]{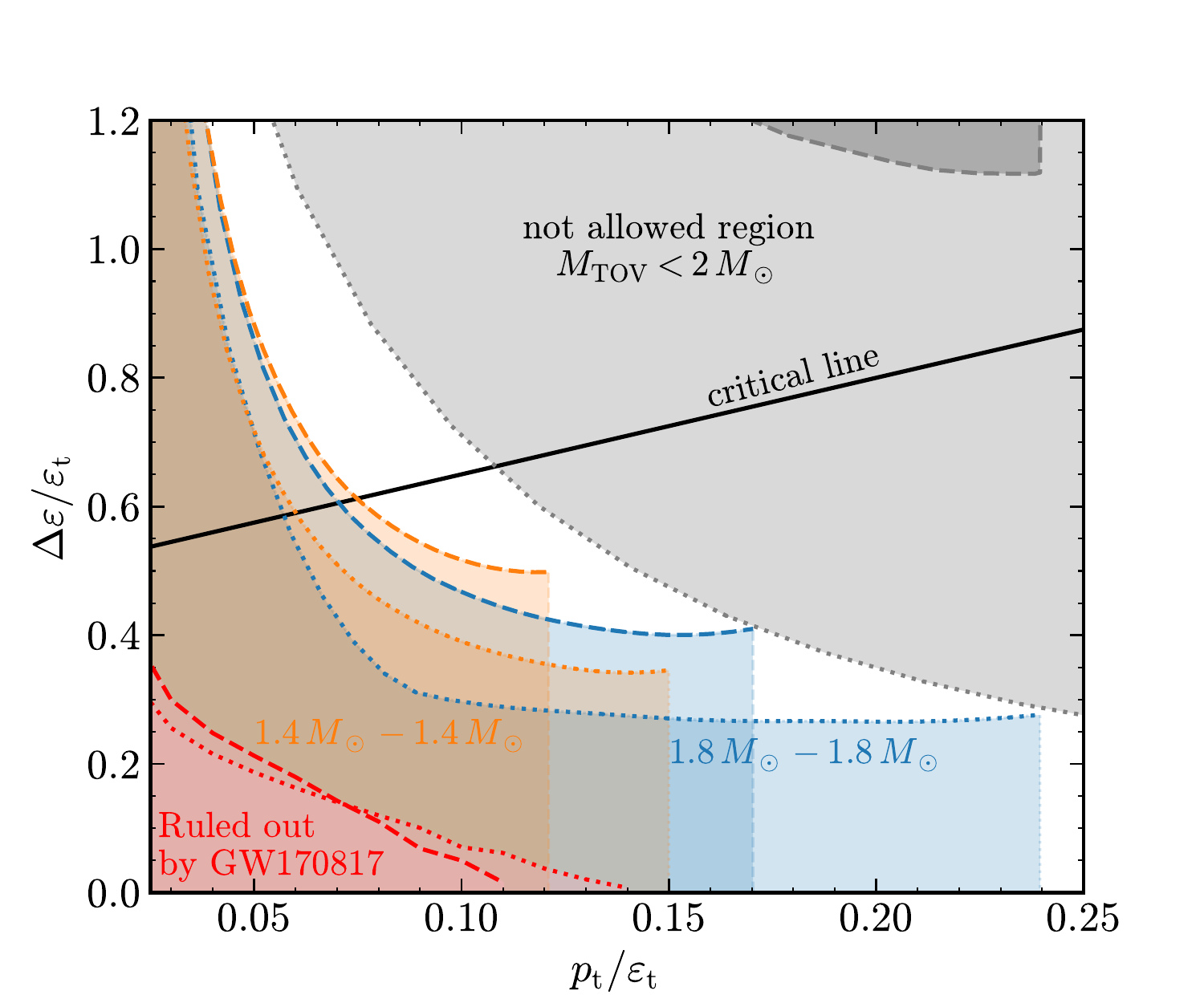} 
    \caption{Detectable regions of phase transition parameters for a $1.4\Msun-1.4\Msun$ system (orange) and a $1.8\Msun-1.8\Msun$ system (blue) on the 3rd generation detectors. The red region depicts the parameter space ruled out by the mode resonance analysis of GW170817 at 95\% confidence level, and the grey regions represent the portions eliminated by the maximum mass of observed pulsars. The results of QMF+CSS and NL3$\omega\rho$+CSS are represented by dotted and dashed lines, respectively. The critical line corresponds to $\Delta\varepsilon/\varepsilon_{\rm t}=1/2+3p_{\rm t}/2\varepsilon_{\rm t}$, dividing the parameter space into two parts. Below the critical line, there is a hybrid star branch connected to the hadronic branch, while above the critical line, there is no connected hybrid star branch~\citep{2013PhRvD..88h3013A}.}  
\label{fig:CSS space with resonace}
\end{figure}

\subsection{Waveform correction}

Throughout the inspiral process, the binary's individual components will be subject to external tidal forces from their respective companions. Upon the tidal driving frequency nearing the g-mode frequency of the NSs, it becomes possible to trigger internal stellar oscillations through resonance, ultimately causing a phase shift in the GW waveform. With the mode frequency and mode eigenfunction obtained in our GR framework, we then are able to estimate the phase shift of the waveform as~\citep{1994MNRAS.270..611L}:
\begin{equation}\label{eq:single phase shift}
    \delta\phi \simeq-0.12\left(\frac{Q}{0.01}\right)^2\left(\frac{600\Hz}{f}\right)^{2}\left(\frac{2q}{1+q}\right)M_{1.4}^{-4}R_{12}^{2}\ ,
\end{equation}
where $q=M^\prime/M$ with $M^\prime$ representing the companion star mass. $M_{1.4}=M/1.4\Msun$ and $R_{12}=R/12\km$. 
The total phase correction for a binary NS waveform $h(f)=\mathcal{A}e^{i\Psi (f)}$, which takes into account the mode resonances of both components, is expressed as \citep{2017MNRAS.470..350Y,2020PhRvL.125t1102P}:
\begin{equation}\label{eq:total phase shift}
\begin{split}
    \Delta \Psi(f) &= -\sum_{i}\delta\phi_i\left(1-\frac{f}{f_i}\right)\Theta(f-f_i)\\ 
    &\approx -\delta\bar\phi\left(1-\frac{f}{\bar f}\right)\Theta(f-\bar f)\ ,
\end{split}
\end{equation}
where $f_i$ is the g-mode frequency of the $i$-th star and $\delta\bar\phi_i$ represents the corresponding phase shift due to resonance. 
Here, $\Theta(\cdot)$ is the Heaviside step function. 
The second line of Eq.~(\ref{eq:total phase shift}) introduces the total phase shift $\delta\bar\phi=\sum_i\delta\phi_i$ and the resonant frequency $\bar f=\delta\bar\phi/\sum_i(\delta\phi_i/f_i)$ to reduce the number of parameters \citep{2020PhRvL.125t1102P}. 
For simplicity, $|\delta\bar\phi|$ will be used as the waveform parameter from here on.

\section{Results}
\subsection{Detectability}

Fig.~\ref{fig:detectability} displays the total phase shift $|\delta\bar\phi|$ against the resonant frequency $\bar f$ for two equal-mass binary systems, along with the detection thresholds of aLIGO and 3rd generation detectors. These thresholds are computed using a Fisher matrix method~\citep{1994PhRvD..49.2658C} and are based on the designed sensitivities~\citep{2017CQGra..34d4001A} assuming that the binary systems are located at $D_{L}=100\,{\rm Mpc}$ (details in Appendix~\ref{appendix: detectability}). Generally, $|\delta\bar\phi|$ decreases with the increase of $\bar f$ since $\delta\phi\propto f^{-2}$ [cf. Eq. (\ref{eq:single phase shift})]~\citep{1994MNRAS.270..611L}. For a significant number of hybrid star models, the total phase shifts are greater than the detectability threshold of 3rd generation detectors.
It is worth noting that there is a threshold frequency $f_{\rm th}$, which is independent of the hadronic EOS. Binaries with $\bar f\lesssim f_{\rm th}$ may be detectable by 3rd generation detectors, while those with $\bar f\gtrsim f_{\rm th}$ may not be detectable. For a $1.4\Msun-1.4\Msun$ system, $f_{\rm th}\simeq 1200\Hz$, while for a $1.8\Msun-1.8\Msun$ system, $f_{\rm th}\simeq 1050\Hz$. These threshold frequencies lead to constraints on the phase transition parameters, which are illustrated in the shaded regions (orange for $1.4\Msun-1.4\Msun$ system and blue for $1.8\Msun-1.8\Msun$ system) in Fig.~\ref{fig:CSS space with resonace}.
It is evident that the detection of a signal of mode resonance provides a new opportunity to probe the properties of the hadron-quark phase transition, which can serve as a complementary constraint to other observations, such as the $\sim2\Msun$ maximum mass constraint from the most massive pulsars (grey regions in Fig.~\ref{fig:CSS space with resonace}). 

 \begin{figure}
  \centering
  \includegraphics[width=0.49\textwidth]{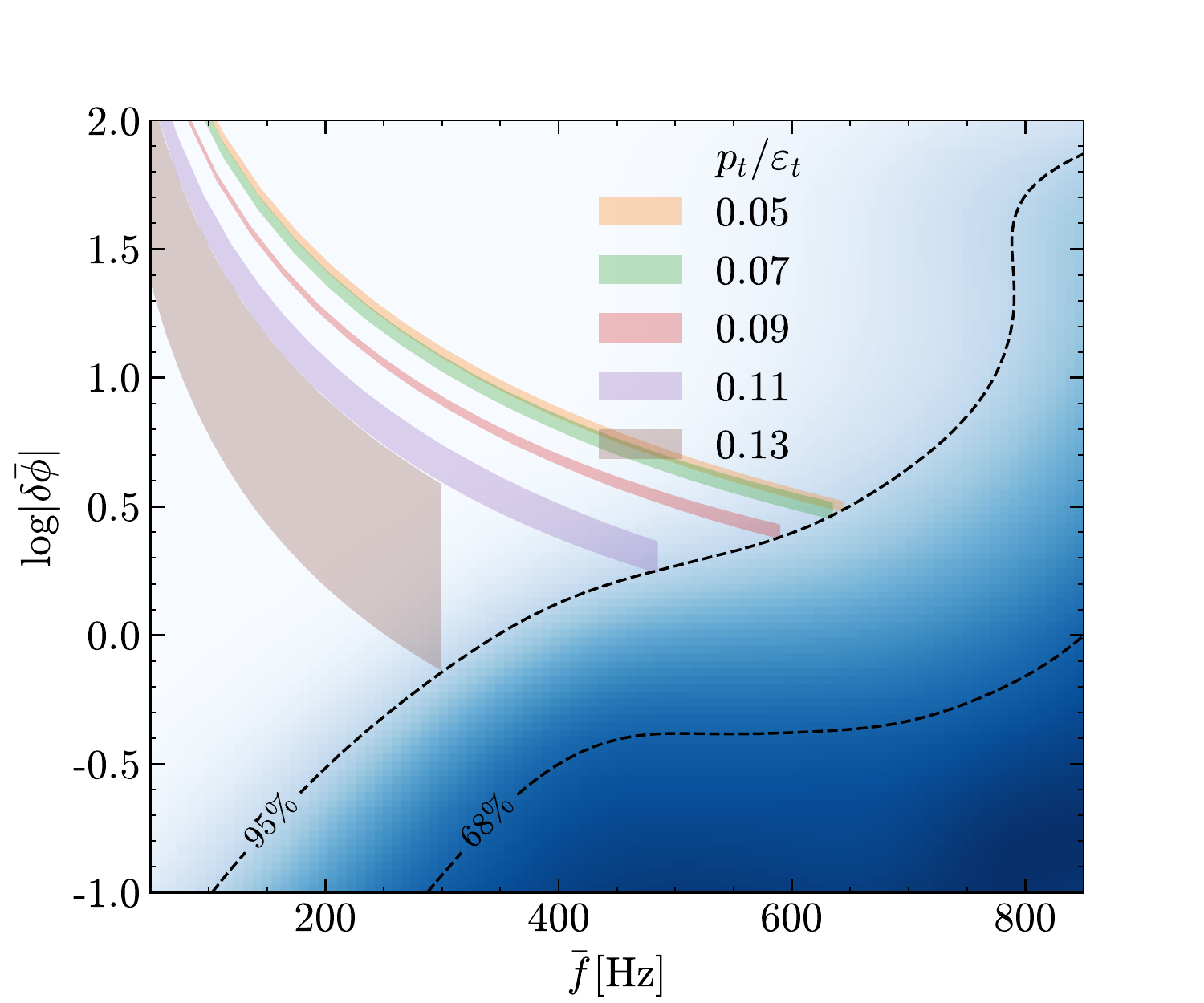}
 \vspace{-0.5em}
  \includegraphics[width=0.49\textwidth]{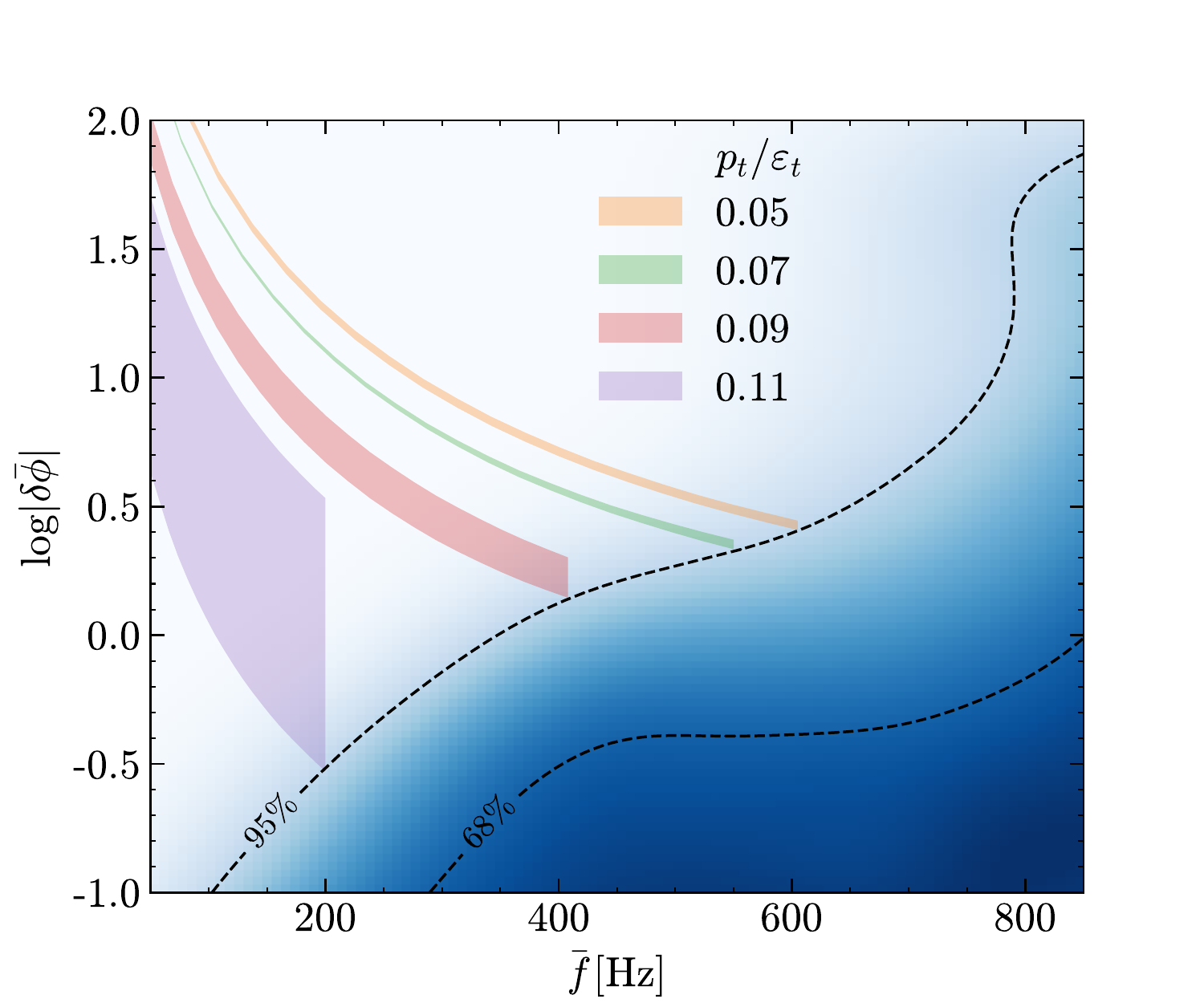}
    \caption{Constraints on phase transition densities (represented by $p_{\rm t}/\varepsilon_{\rm t}$) from GW170817 data for QMF+CSS (upper panel) and NL3$\omega\rho$+CSS (lower panel). Colored bands depict calculations for a GW170817-like system with chirp mass $\mathcal{M}=1.186\Msun$ and mass ratio $q=0.7$--$1$. 
     The posterior probability density for $\Delta|\bar\phi|$ and $\bar f$ inferred from the GW170817 data is shown in blue heatmap, with the 95\% and 68\% boundaries being overlaid by dashed lines. 
     } 
\label{fig:GW170817 system}
\end{figure}

\subsection{GW170817 constraints}

We proceed by making use of the available GW170817 data and conduct a search for mode resonance signals in full range of plausible resonant frequencies $[50,1600]\Hz$ (the upper limit corresponds to the shut-off frequency of the waveform, $4400\Hz/(M+M^\prime)\approx 1600\Hz$, for the GW170817 system)\footnote{In principle, the g-mode frequency can be made arbitrarily small if $\Delta\varepsilon$ is sufficiently tiny. However, given the substantial noise levels at low frequencies ($\lesssim10\Hz$), we have opted for a lower limit of $50\Hz$ for the search. This choice corresponds to a transition strength of $\Delta\varepsilon\simeq 1\,{\rm MeV}$.}, with the application of waveform correction as described in Eq.~(\ref{eq:total phase shift}). 
This is similar to a previous search conducted in \citet{2020PhRvL.125t1102P} for signals in the frequency range of $[30,300]\Hz$. 
We employ a Bayesian parameter estimation approach, following the data analysis framework presented in \citet{2019PhRvX...9a1001A}. 
Our analysis utilizes {\sc PyMultiNest}~\citep{2014A&A...564A.125B} implemented in {\sc Bilby}~\citep{2019ApJS..241...27A}, enabling us to simultaneously obtain the posterior distribution of parameters and the model evidence $\mathcal{Z}$. 
By denoting $\mathcal{H}_1$ as the hypothesis with mode resonance and $\mathcal{H}_0$ as the one without, we are able to compute the Bayes factor from the evidence, $\mathcal{B}^1_0=\mathcal{Z}_1/\mathcal{Z}_0$.
We obtain a Bayes factor of $\mathcal{B}^1_0=0.72$, which implies that the current data does not strongly favor one hypothesis over the other (see Appendix~\ref{appendix: bayesian analysis} for more discussions on the Bayesian parameter estimation). 

Though the search yields no results, the posterior distribution of $\Delta|\bar\phi|$ and $\bar f$ can be utilized to set constraints on the phase transition parameters. 
In Fig.~\ref{fig:GW170817 system} we present the calculated $\Delta|\bar\phi|$ and $\bar f$ for a GW170817-like system with a chirp mass of $\mathcal{M}=1.186\Msun$ and a mass ratio range of $q=0.7$--$1$. 
The posterior distribution for $\Delta|\bar\phi|$ and $\bar f$ together with their 95\% and 68\% contours are also plotted. 
It can be seen that for a certain transition density, the hybrid star models with small resonant frequencies should be excluded. For example, those models with $p_{\rm t}/\varepsilon_{\rm t}=0.11$ and $\bar f\leq 485\Hz$ are ruled out at 95\% confidence level for QMF+CSS.
Converting these constraints on $\bar{f}$ to phase transition properties, we conclusively exclude the possibility of a weak phase transition occurring at low densities with a confidence level of 95\%. 
This exclusion is represented in Fig.~\ref{fig:CSS space with resonace} by the red region.
Our conclusion here is in agreement with our previous findings based on static tidal deformability constraints from GW170817~\citep{2020ApJ...904..103M} and the study of proton and lambda production during relativistic heavy-ion collisions at several GeV/nucleon incident beam energies~\citep{2023PhRvD.107d3005L}.

\section{Discussion}
The investigation of phase transitions in QCD matter is a central area of interest in physics and has relevance for a range of open problems, such as early Universe~\citep{1983PhRvL..51.1488H,2022PhRvD.106l1902C}, core-collapse supernovae~\citep{2018NatAs...2..980F}, cosmological gamma-ray bursts~\citep{1996PhRvL..77.1210C}, as well as the nonperturbative properties of QCD itself~\citep{2020PhR...879....1H}.
We propose to probe the phase transition properties possible in dense NS matter through the tidal resonant excitation of g-modes. 

Our analysis demonstrates that the tidal excitation of g-modes from inspiralling neutron stars can induce significant phase shifts in the gravitational waveform, on the order of $\sim\mathcal{O}(0.1)-\mathcal{O}(1)$. These shifts are detectable by both aLIGO and 3rd generation detectors. Notably, these low-frequency signatures complement previous efforts to identify first-order phase transitions using high-frequency post-merger gravitational waves~\citep{2019PhRvL.122f1102B}, expanding the range of density regimes that can be probed for potential phase transitions. 
We provide two analytical formulas to clarify how the phase shift in the gravitational waveform relates to the properties of the phase transition. These formulas will serve as valuable tools for future attempts to directly estimate phase transition parameters using GW data. 
Our analysis suggests that the estimation errors are moderate, with a maximum of $\sim15\%$ for the mode frequencies and 25\% for the tidal overlap integrals. This implies a potential underestimation or overestimation of the phase shift by no more than a factor of 2 to 3, as detailed in the Appendix~\ref{appendix: approx frequency}-\ref{appendix: approx tidal overlap}. 
Even in cases where no mode resonance signal is detected, the search results can effectively rule out a significant parameter space, as demonstrated earlier with GW170817.

To calculate the correction to the gravitational waveform induced by the g-mode resonance, we use the form presented in Eq.~(\ref{eq:total phase shift}). 
However, it is important to note that this form is derived under the assumption that the energy transfer of the resonance is instantaneous~\citep{2017MNRAS.470..350Y}. Therefore, it is only valid when the duration of the resonance, $t_{\rm res}\simeq 0.009s\,(\mathcal{M}/1.2\Msun)^{-5/6}(f/600\,{\rm Hz})^{-11/6}$, is much shorter than the orbital decay time-scale, $t_D \simeq 0.08s\,(\mathcal{M}/1.2\Msun)^{-5/3}(f/600\Hz)^{-8/3}$~\citep{1994MNRAS.270..611L}. For resonant frequencies that are very high (e.g., $\gtrsim1000\Hz$), the waveform correction $\Delta\Psi(f)$ should be modified to account for the continuous process of energy transfer during the resonance. 
It is also important to consider the relativistic corrections to the Newtonian orbit as the binary approaches closer, especially for high resonant frequencies. 
As demonstrated by \citet{1992ApJ...400..175B}, tidal interaction is not sufficient to spin up the binary NS system before the merger and an ir-rotational merger is expected in most cases due to the spinning-down by magnetic dipole radiation before the merger. At present, we do not take into consideration the star rotation. Nevertheless, NS spin could be important in cases such as dynamic captures, which can shift the mode frequency to lower values and thus can contribute to larger GW phase shift~\citep{1999MNRAS.308..153H}. 
These improvements will be the subject of future work. 

In addition, it is crucial to identify the physical origin of the mode resonance signature once it is detected. This is because there could be other possible origins apart from the resonant excitation of hybrid star g-modes. As suggested by \citet{2020PhRvL.125t1102P}, other processes such as tidal-p-g coupling~\citep{2013ApJ...769..121W} or those in modified gravity~\citep{2014PhRvD..89d4024P,2019PhRvD..99d4038A,2021PhRvD.103l4034K,2021PhRvD.104j4036M} predict different corrections on the waveform. Therefore, one could perform a Bayesian model selection to compare these different origins. 
On the other hand, crust-core i-mode or core composition g-mode can produce a $\Delta\Psi(f)$ similar to that studied in this work. For the former, its resonant frequency is typically around $100\Hz$, while for the latter, its frequency can reach up to several hundred Hz according to recent studies using the Cowling approximation~(see e.g.~\citet{2021PhRvD.103l3009J}).
Our Fisher analysis shows that the resonant frequencies can be measured with a precision of $\lesssim$10\% using 3rd generation detectors (see Appendix~\ref{appendix: detectability}). This precision would enable us to discriminate between the discontinuity g-mode resonance considered here and the crust-core i-mode or composition g-mode by comparing their frequencies.
It is worth mentioning that the existence and properties of QCD phase transition are still very unclear in nuclear physics studies. Various scenarios, including cross-over/first-order transitions and fast/slow transitions, lack effective constraints. The analysis of this paper is based on the slow first-order phase transition scenarios. We might expect future GW observations of binary NS mergers could exclude or put constraints on such scenarios, hence contributing to our understanding of the QCD phase diagram.

\appendix

\subsection{Solving pulsation equations in general relativity}\label{appendix: pulsation equation}
For the background star, the static and spherically symmetric line element is given by
\begin{equation}
	ds^2 =g_{\mu\nu}dx^\mu dx^\nu= -e^{2\Phi}dt^2+e^{2\lambda}dr^2+r^2(d\theta^2+\sin^2\theta d\phi^2)
\end{equation}
where the metric components $\Phi$ and $\lambda$ are functions of $r$ only. We model the star as a perfect fluid, with the stress-energy-momentum tensor given by
\begin{equation}
    T_{\mu\nu} = (p+\varepsilon)u_\mu u_\nu+pg_{\mu\nu},
\end{equation}
where $u^\mu$ is the four-velocity. We also assume the metric perturbation as
\begin{equation}\label{eq:metric pertur}
    h_{\mu\nu} = -r^lH_0e^{i\omega t}Y_{lm}dt^2-r^lH_0e^{i\omega t}Y_{lm}dr^2-r^lKe^{i\omega t}Y_{lm}r^2(d\theta^2+\sin^2\theta d\phi^2)-2i\omega r^{l+1}H_1e^{i\omega t}Y_{lm}dtdr,
\end{equation}
and the Lagrangian displacement as
\begin{equation}\label{eq:displacement}
\begin{split}
    \xi^r = &r^{l-1}e^{-\lambda}W Y_{lm}e^{i\omega t},\\
	\xi^\theta = &-r^{l-2}V \partial_\theta Y_{lm}e^{i\omega t},\\
	\xi^\phi = &-r^{l-2}\sin^{-2}\theta V\partial_\phi Y_{lm}e^{i\omega t},
\end{split}
\end{equation}
where $Y_{lm}(\theta,\phi)$ is the spherical harmonics. Note the Eulerian perturbations of energy density and pressure are given by
\begin{equation}\label{eq:pe pertur}
\begin{array}{ll}
    \delta \varepsilon &= (p+\varepsilon)(\Delta n/n)-\xi^i\frac{d\varepsilon}{dx^i},\\
    \delta p &= \gamma p(\Delta n/n)-\xi^i\frac{dp}{dx^i}.
\end{array}
\end{equation}
where $\gamma = [(p+\varepsilon)/p](\partial p/\partial \varepsilon)_{\rm ad}$ is the adiabatic index. The Lagrangian perturbation of baryon number density $\Delta n$ can be derived from the baryon number conservation $\nabla_\mu(nu^\mu)=0$, it reads
\begin{equation}
    \Delta n = -n\nabla_i\xi^i-\frac12n\delta g^{(3)}/g^{(3)},
\end{equation}
with $g^{(3)}$ the determinant of the metric of the 3-geometry at constant time.
Plugging these perturbations (Eqs.~\ref{eq:metric pertur}-\ref{eq:pe pertur}) into linear Einstein equation $\delta G_{\mu\nu}=8\pi\delta T_{\mu\nu}$, one finds~\citep{1983ApJS...53...73L,1985ApJ...292...12D}
\begin{align}
    \frac{dH_1}{dr}=&-\bigg[l+1+\frac{2m}{r}e^{2\lambda}+4\pi r^2(p-\varepsilon)e^{2\lambda}\bigg]\frac{H_1}{r}+\frac{e^{2\lambda}}{r}\bigg[H+K-16(\varepsilon+p)V\bigg],\label{eq:H1}\\
    \frac{dK}{dr}=&\frac{l(l+1)}{2r}H_1+\frac{H_0}{r}-\left(l+1-r\Phi^\prime\right)\frac{K}{r}-\frac{8\pi}{r}(\varepsilon+p)e^\lambda W,\label{eq:K}\\
    \frac{dW}{dr} = &-\frac{l+1}{r}W+re^\lambda\bigg[\frac{1}{\gamma p}e^{-\Phi}X-\frac{l(l+1)}{r^2}V+\frac12H_0+K\bigg],\label{eq:W}\\
    \frac{dX}{dr}=&-\frac{l}{r}X+(\varepsilon+p)\frac{e^\Phi}{r}\Bigg\{\frac12(1-r\Phi^\prime)H_0+\frac12\left(\omega^2r^2e^{-2\lambda}+\frac{l(l+1)}{2}\right)H_1-\left(\frac{1}{2}-\frac32r\Phi^\prime\right)K-\frac{l(l+1)}{r}\Phi^\prime V\nonumber\\
    &-\bigg(\omega^2e^{\lambda-2\Phi}+4\pi(\varepsilon+p)e^\lambda-r^2(\frac{1}{r^2}e^{-\lambda}\Phi^\prime)^\prime\bigg)W\Bigg\}, \label{eq:X}
\end{align}
with 
\begin{equation}
	X \equiv \omega^2(p+\varepsilon)e^{-\Phi}V-r^{-1}e^{\Phi-\lambda}p^\prime W+\frac12(p+\varepsilon)e^\Phi H_0.
\end{equation}
and 
\begin{equation}
\begin{split}
	(\frac{3m}{r}+\frac{(l+2)(l-1)}{2}+4\pi r^2p)H_0 = &8\pi r^2e^{-\Phi}X+r^2e^{-2\lambda}\bigg[\omega^2e^{-2\Phi}-\frac{l(l+1)}{2}\Phi^\prime\bigg]H_1\\
	&+\bigg[\frac{(l+2)(l-1)}{2}-\omega^2r^2e^{-2\Phi}+(r-3m-4\pi r^3p)\Phi^\prime\bigg]K.
\end{split}
\end{equation}
Here the prime denotes the derivative of $r$. 

\begin{figure}[ht]
    \centering
    \includegraphics[width=3.2in]{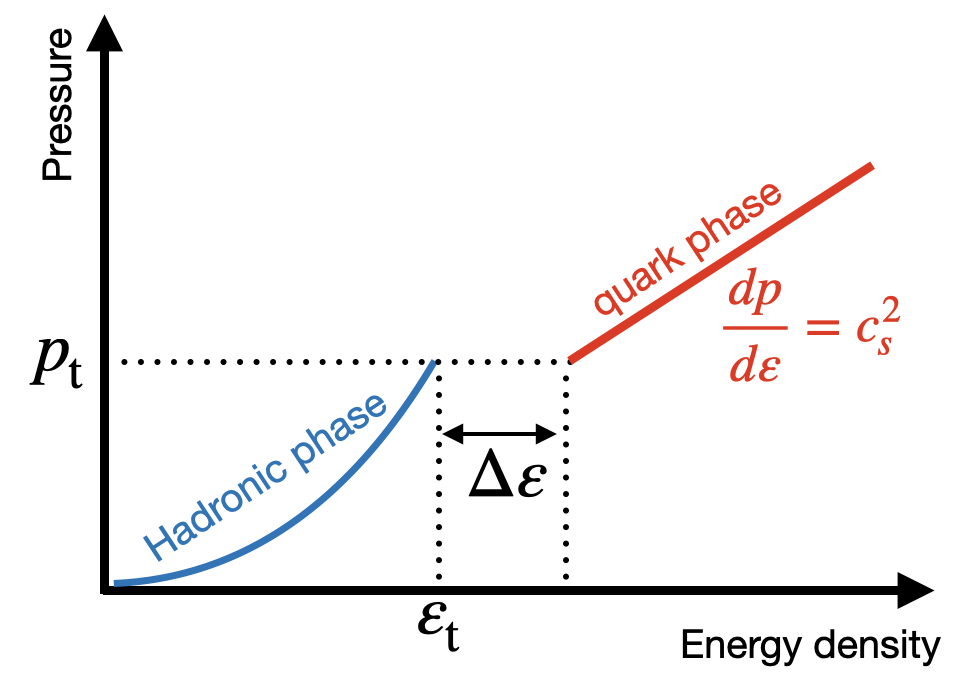}
    \caption{An illustration of our NS EOS with a first-order phase transition to quark matter.} 
\label{fig:CSS EOS}
\end{figure}

\begin{figure}[ht]
    \centering
  \begin{minipage}{0.48\linewidth}
    \centering
    \includegraphics[width=3.2in]{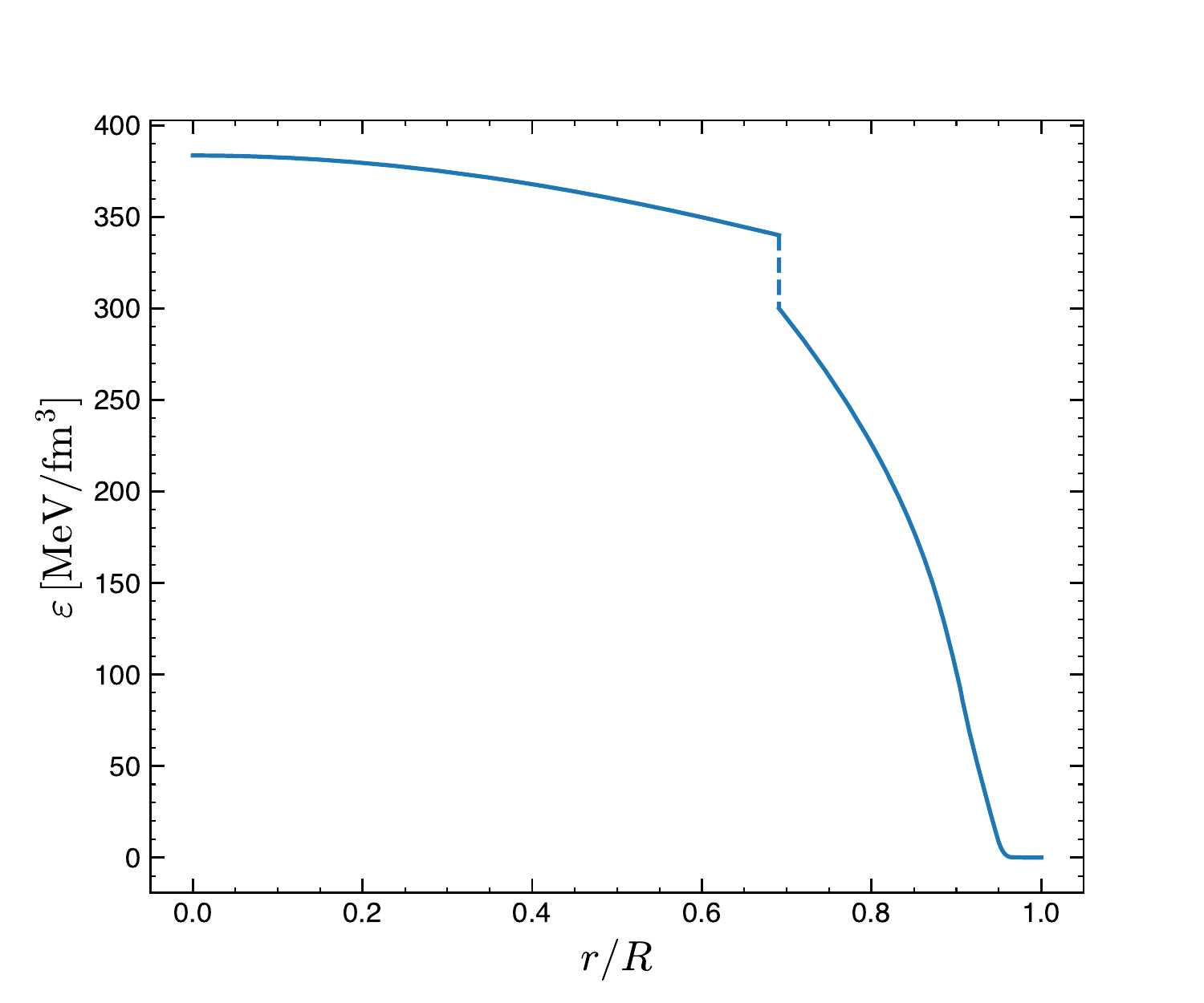}
  \end{minipage}
    \begin{minipage}{0.48\linewidth}
    \centering
    \includegraphics[width=3.2in]{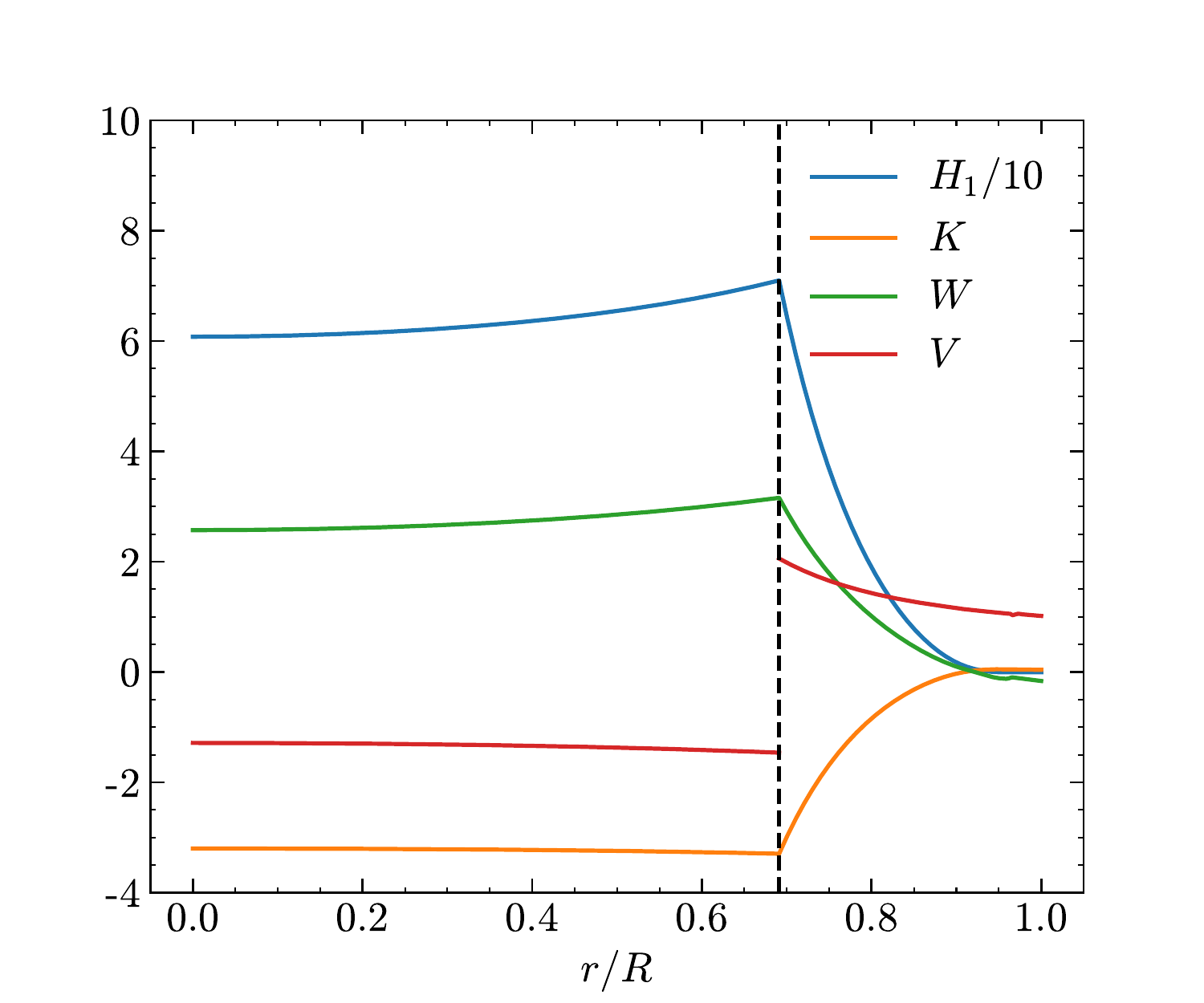}
  \end{minipage}
    \caption{Left: Density profile for a $1.4\Msun$ hybrid star, with $\varepsilon_{\rm t}=300\,{\rm MeV/fm^3}$, $\Delta\varepsilon=40\,{\rm MeV/fm^3}$ and $c_s=1$ within QMF+CSS; Right: Perturbation functions of the discontinuity g-mode corresponding to the star shown in the left panel. The unit of metric perturbation in the right panel is $153\,{\rm MeV/fm^3}$.} 
\label{fig:example solution}
\end{figure}

Quark matter is present from the first-order transition onset up to the maximum central pressure of a star (see Fig.~\ref{fig:CSS EOS}). Our description of the EOS is written as follows:
\begin{equation}
\varepsilon(p) = \left\{\!
\begin{aligned}
&\varepsilon_{\rm HM}(p), & p<p_{\rm t} \\ 
&\varepsilon_{\rm HM}(p_{\rm t})+\Delta\varepsilon+c_s^{-2} (p-p_{\rm t}), & p>p_{\rm t}  
\end{aligned}
\right.
\end{equation}
where `HM' denotes hadronic matter.
We then follow the method described in \citet{1983ApJS...53...73L} to solve these equations and find the eigenvalue of $\omega$ for each mode. 
It's particularly important to note that when solving the aforementioned perturbations equations, we require the boundary condition that $\Delta p=-r^le^{-\Phi}XY_{lm}e^{i\omega t}$ to be continuous at the hadron-quark interface. This necessitates that at the transition radius $R_{\rm t}$,
\begin{equation}
    V_+ = \frac{\varepsilon_{\rm t}+\Delta\varepsilon+p}{\varepsilon_{\rm t}+p}V_-+\frac{\Delta\varepsilon}{\omega^2(\varepsilon_{\rm t}+p)}\left(\frac{e^{2\Phi-\lambda}}{R_{\rm t}}\Phi^\prime W+\frac12e^{2\Phi}H_0\right)
\end{equation}
where the subscripts `$+$' and `$-$' denote the values at $R_{\rm t}+0^+$ and $R_{\rm t}-0^+$, respectively. This implies that at both ends of the interface, although the radial displacement W is continuous, the tangential displacement V is discontinuous due to the presence of a density discontinuity. It is this particularity that gives rise to the discontinuity g-mode studied in this work. 
Fig.~\ref{fig:example solution} provides a solved example, allowing us to observe the density jump and the corresponding variation of $V$ at the interface.

\subsection{Relativistic tidal overlap integral}\label{appendix: tidal overlap}

\begin{table}
\centering
\caption{Comparison for the numerical tidal overlap $Q$ and g-mode overlap $\langle \xi_g|\xi_\alpha \rangle$ obtained by using different approaches. The EOS model is QMF+CSS with $\varepsilon_{\rm t}=300\,{\rm MeV/fm^3}$ and $\Delta\varepsilon=40\,{\rm MeV/fm^3}$. 
}
\renewcommand\arraystretch{1.5}
\begin{ruledtabular}
\begin{tabular}{ccccccc}
\multirow{3}{*}{$\alpha$} & \multicolumn{2}{c}{\multirow{2}{*}{Hybrid}} & \multicolumn{2}{c}{GR} & \multicolumn{2}{c}{GR}  \\
& & & \multicolumn{2}{c}{(Newtonian inner product)} & \multicolumn{2}{c}{(relativistic inner product)}\\
\cline{2-3}\cline{4-5}\cline{6-7}
 & $\langle \xi_g|\xi_\alpha\rangle$&$|Q_\alpha^N|$ & $\langle \xi_g|\xi_\alpha\rangle$ &$|Q_\alpha^N|$ & $\langle \xi_g|\xi_\alpha\rangle$ &$|Q_\alpha|$\\
\hline
g & 1 & 3.73e-2& 1 & 4.17e-2 & 1&3.45e-2 \\
f & 7.68e-2 & 5.92e-1 &9.79e-3 & 5.86e-1& 7.05e-5&7.08e-1 \\
p$_1$ & -2.71e-2 &1.17e-1 & 3.95e-3 & 4.22e-2 &1.65e-5 & 1.01e-2\\
p$_2$ & 2.60e-2 &5.25e-2& -1.41e-3 & 8.88e-3 &-2.50e-5 & 2.69e-2\\
p$_3$ & -1.82e-2 &2.12e-3 & 2.21e-3 & 1.29e-2 &-2.95e-5 & 4.00e-2\\
\hline
$\sum_\alpha Q_\alpha^2$& \multicolumn{2}{c}{0.3680} & \multicolumn{2}{c}{0.3073}& \multicolumn{2}{c}{0.5150}\\
$\langle \nabla (r^lY_{lm})|\nabla (r^lY_{lm})\rangle$ & \multicolumn{2}{c}{0.3504}& \multicolumn{2}{c}{0.3504}& \multicolumn{2}{c}{0.5175}

\end{tabular}
\end{ruledtabular}
    \vspace{-0.4cm}
\label{table:Q rel}
\end{table}

In some previous works, the hybrid (GR background + Newtonian perturbation) approach was typically used to calculate oscillation modes and the Newtonian formalism was employed to compute tidal overlap integral, i.e., $Q^{N}_\alpha\equiv \langle \xi_\alpha|\nabla (r^lY_{lm})\rangle/\langle \xi_\alpha | \xi_\alpha \rangle$ with the inner product $\langle \xi|\psi\rangle = \int \rho \boldsymbol{\xi}^*\cdot \boldsymbol{\psi} d^3x$. This hybrid approach is inconsistent and will result in computed eigenfunctions that are not orthogonal~\citep{1994ApJ...432..296R}, means $\langle \xi_\alpha|\xi_\beta\rangle \neq 0$ when $\alpha \neq \beta$, see Table~\ref{table:Q rel}. In this case, the obtained $Q^N$ for g-modes would contain some f-mode contamination, leading to unreliable values of $Q^N$.

In this work, we calculate the stellar oscillation modes within the GR framework. This naturally ensures that the eigenfunctions are orthogonal under a relativistic extension of inner product~\footnote[1]{Strictly speaking, the definition of this inner product should also include the integration of metric perturbations out to infinity~\citep{1973ApJ...185..685D}. However, this term is small for g-modes considered in this work and we omit it here.}
\begin{equation}
    \langle \xi^\mu|\psi^\mu\rangle = \int \sqrt{-g}e^{-2\Phi}(p+\varepsilon)\xi^{*\mu}\psi_\mu d^3x.
\end{equation}
Based on this inner product, we find that the relativistic tidal overlap integral can still be defined as 
\begin{equation}
    Q_\alpha = \frac{\langle \xi_\alpha|\nabla (r^lY_{lm})\rangle}{\langle \xi_\alpha | \xi_\alpha \rangle}.
\end{equation}
In the following, we give a brief derivation and test the accuracy of our numerical results.

We begin from the orthogonal part of the Euler equation $\nabla_\nu T^{\mu\nu}=0$, which is expressed as 
\begin{equation}\label{eq:euler orth}
    (p+\varepsilon)u^\nu\nabla_\nu u^\mu +\perp^{\mu\nu}\nabla_\nu p=0,
\end{equation}
where $\perp^{\mu\nu}=g^{\mu\nu}+u^\mu u^\nu$ is the projection operator. Plugging the the Lagrangian displacement of a certain mode $\alpha$, $\xi_\alpha^\mu=(0,\boldsymbol{\xi}_\alpha(r)e^{i\omega_\alpha t})$, into Eq.~(\ref{eq:euler orth}) one can derive the linear pulsation equation
\begin{equation}\label{eq:eigenval omega}
    [\mathcal{L}-(p+\varepsilon)e^{-2\Phi}\omega_\alpha^2]\xi_\alpha^\mu = 0,
\end{equation}
with 
\begin{equation}
    \mathcal{L}\xi_\alpha^\mu = (\delta p+\delta \varepsilon) u^\nu\nabla_\nu u^\mu+\perp^{\mu\nu}\nabla_\nu\delta p+i\omega_\alpha e^{-\Phi}\left[(p+\varepsilon)\nabla_\nu u^\mu+u^\mu\nabla_\nu p\right]\xi_\alpha^\nu+i\omega_\alpha e^{-\Phi}\xi_\alpha^\mu u^\nu\nabla_\nu p.  
\end{equation}

During the inspiral, the linear perturbation of the tidal potential is governed by
\begin{equation}\label{eq:linear eq}
    (p+\varepsilon)u^\nu\partial_\nu\delta u^\mu +(p+\varepsilon) u^\nu\delta\Gamma^{\mu}_{\nu\rho}u^\rho+\mathcal{L}\xi^\mu = 0.
\end{equation}
Plugging Eq.~(\ref{eq:eigenval omega}), $u^{\mu} =(e^{-\Phi},0,0,0)$, $\delta u^i = e^{-\Phi}\partial_t\xi^{i}$ and $\delta\Gamma^\mu_{00}=-1/2g^{\mu\rho}\partial_\rho h^{\rm tid}_{00}$ into Eq.~(\ref{eq:linear eq}), one obtains
\begin{equation}\label{eq:eq of motion}
    (p+\varepsilon)e^{-2\Phi}\frac{\partial^2}{\partial t^2}\xi^\mu + \mathcal{L}\xi^\mu = -(p+\varepsilon)e^{-2\Phi}\partial^\mu U,
\end{equation}
where $U=-1/2h^{\rm tid}_{00} = -M^\prime /|\boldsymbol{r}-\boldsymbol{D}|$ is the tidal potential, $D$ is the distance of the secondary component $M^\prime$. 
Eq.~(\ref{eq:eq of motion}) is similar to its Newtonian counterpart, as seen, for instance, in Eq.(2.5) of \citet{1994MNRAS.270..611L}. Hence, similar to \citet{1994MNRAS.270..611L}, we arrive to the relativistic definition of tidal overlap integral as
$Q_\alpha = \langle \xi_\alpha|\nabla (r^lY_{lm})\rangle/\langle \xi_\alpha | \xi_\alpha \rangle$.
In numerical calculation, we choose the normalization $\langle \xi_\alpha|\xi_\alpha\rangle = MR^2$ for convenience.

To test the accuracy of our numerical results, we first calculate the inner product $\langle \xi_g|\xi_\alpha\rangle$, the exemplary results are shown in Table~\ref{table:Q rel}. It can be seen that for $\alpha\neq g$, the inner products we obtained are very tiny, which verifies the orthogonality of the eigenfunctions. We also check the sum rule of $Q_\alpha$, namely $\sum_\alpha Q_\alpha^2=\langle \nabla (r^lY_{lm})|\nabla (r^lY_{lm})\rangle$, which is a general property of the tidal overlap integral~\citep{1994ApJ...432..296R}. From Table~\ref{table:Q rel} we see that our numerical $Q$ obeys the sum rule.

\subsection{Approximated formulas of the mode frequency}\label{appendix: approx frequency}
Previously, \citet{1987MNRAS.227..265F} pointed out that when the discontinuity near the surface (i.e., $R-R_{\rm t}\ll R_{\rm t}$), the frequency can be approximated as
\begin{equation}\label{eq:Finn}
    \omega^2 \approx l(l+1) \frac{\Delta\varepsilon}{\varepsilon_{\rm t}+\Delta\varepsilon}\frac{R-R_{\rm t}}{R} \frac{M}{R^3}.
\end{equation}
And this approximation has been improved by \citet{2022PhRvD.106l3002Z} with
\begin{equation}\label{eq:Zhao}
    \omega^2 \approx \frac{D \Delta\varepsilon/\varepsilon_{\rm t}}{(1+\Delta\varepsilon/\varepsilon_{\rm t})/\tanh [D]+1/\tanh [D(R/R_{\rm t}-1)]}\frac{M_{\rm t}}{R_{\rm t}^3}
\end{equation}
where $M_{\rm t}$ is the mass inside $R_{\rm t}$ and $D=1.21$ is a fitting coefficient.

\begin{figure}[ht]
    \centering
    \includegraphics[width=2.9in]{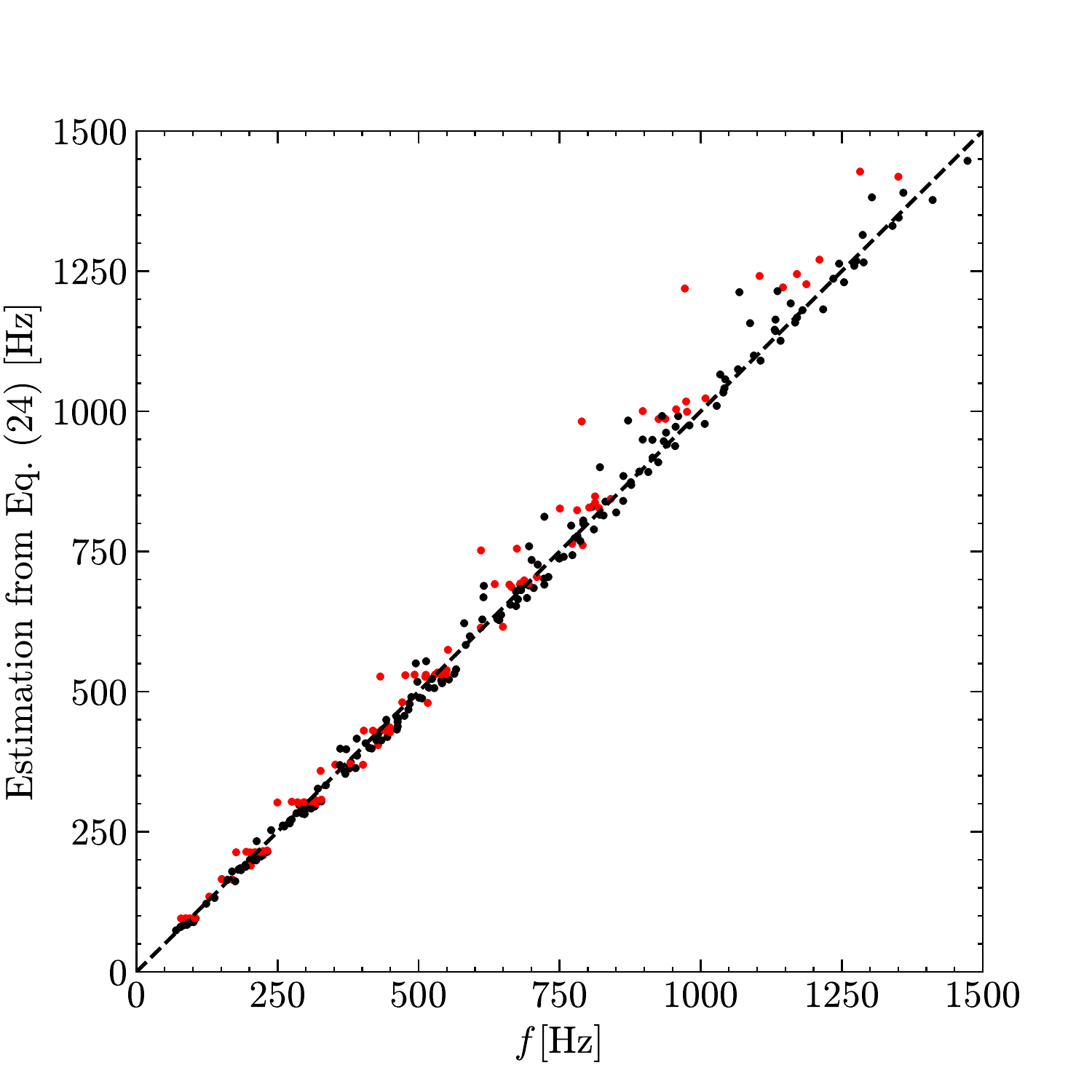}
    \caption{Estimated frequencies from Eq.~(\ref{eq:omega fit}) versus the frequencies calculated numerically from relativistic pulsation equations.} 
\label{fig:f_estimation}
\end{figure}

However, both approximations of \citet{1987MNRAS.227..265F} and \citet{2022PhRvD.106l3002Z} start from the analytical solution of the frequency of surface gravity waves between two incompressible fluids, obtained by \citet{1959flme.book.....L} assuming the slab configuration. Thus it should be noted that their approximations can not seamlessly cater to both the $R-R_{\rm t}\ll R$ and $R_{\rm t}\ll R$ cases, see Fig. 13 in \citet{2022PhRvD.106l3002Z}. 
In this work, we present a more accurate approximation that can accommodate both these two cases simultaneously, i.e.,
\begin{equation}\label{eq:omega fit}
    \omega^2 \approx K\frac{(6/5)(\Delta\varepsilon/\varepsilon_{\rm t})(1-x^5)+1+\varsigma-\sqrt{(1+\varsigma)^2-(12/5)(\Delta\varepsilon/\varepsilon_{\rm t})(\kappa-1)(1-x^5)}}{(\varepsilon_{\rm t}+\Delta\varepsilon)/\varepsilon_{\rm t}-(2/5)(\Delta\varepsilon/\varepsilon_{\rm t})(1-x^5)}\frac{M}{R^3},
\end{equation}
where $\varsigma = (1/5)(\Delta\varepsilon/\varepsilon_{\rm t})[3(\kappa-1)+(3+2\kappa)x^5]$ with $x=R_{\rm t}/R$ and $\kappa=(M_t/M)(R/R_{\rm t})^3$. $K=0.59$ is a fitting coefficient. Fig.~\ref{fig:f_estimation} shows the approximation of Eq.~(\ref{eq:omega fit}). It can be seen that the estimation of Eq.~(\ref{eq:omega fit}) performs well for both $R_{\rm t}<R/2$ and $R_{\rm t}>R/2$ regions. The estimation errors of Eq.~(\ref{eq:omega fit}) are found to be not more than $\sim15\%$.

We should emphasize here that, apart from a factor $K$, Eq.~(\ref{eq:omega fit}) is the exact solution for a two-layer fluid in a spherical configuration within the Newtonian limit. In the following we present a brief derivation.

We begin with the pulsation equation under the Cowling approximation~\citep{1941MNRAS.101..367C}
\begin{equation}\label{eq:NT pert}
\begin{split}
    \frac{d\xi_r}{dr}&=\left(\frac{gr}{c_{ad}^2}-2\right)\frac{\xi_r}{r}+\left[l(l+1)-\frac{\omega^2r^2}{c_{ad}^2}\right]\frac{\xi_h}{r},\\
    \frac{d\xi_h}{dr}&=\left(1-\frac{N^2}{\omega^2}\right)\frac{\xi_r}{r}+\left(\frac{N^2r}{g}-1\right)\frac{\xi_h}{r},
\end{split}
\end{equation}
where $\xi_r$ and $\xi_h$ are the radial and tangential displacement, respectively. 
$g$ is the gravitational acceleration. $N^2=g^2(1/c_e^2-1/c_{ad}^2)$ is the Brunt-V\"ais\"al\"a frequency with $c_e$ and $c_{ad}$ the equilibrium and adiabatic sound speed, respectively. For simplicity, we shall omit the gravity and composition gradient, i.e., we assume $c_{ad}=c_e$, $gr/c_{ad}^2=0$ and $\omega^2r^2/c_{ad}^2=0$. The density for the lower layer (i.e., quark core) and upper layer (i.e., hadronic envelope) are denoted as $\rho^-$ and $\rho^+$. 
Restrict to $l=2$, one finds that Eq.~(\ref{eq:NT pert}) reduces to 
\begin{equation}\label{eq:NT pert reduced}
\begin{split}
    \frac{d\xi_r}{dr}&=-\frac{2\xi_r}{r}+\frac{6\xi_h}{r},\\
    \frac{d\xi_h}{dr}&=\frac{\xi_r}{r}-\frac{\xi_h}{r},
\end{split}
\end{equation}
The general solutions can be obtained with
\begin{equation}\label{eq:xi_r}
\xi_r = \left\{\!
\begin{aligned}
&r,  & 0<r<R_{\rm t}, \\ 
& \frac{\alpha R_{\rm t}^4}{r^4}+\frac{\beta r}{R_{\rm t}}, & R_{\rm t}<r<R.
\end{aligned}
\right.
\end{equation}
and 
\begin{equation}\label{eq:xi_h}
\xi_h = \left\{\!
\begin{aligned}
&\frac12r,  & 0<r<R_{\rm t}, \\ 
& -\frac13\frac{\alpha R_{\rm t}^4}{r^4}+\frac12\frac{\beta r}{R_{\rm t}}, & R_{\rm t}<r<R.
\end{aligned}
\right.
\end{equation}
At the interface of two layers, $\xi_r$ and $\Delta p = \rho g(\omega^2r\xi_h-\xi_r)$ should be continuous, which requires
\begin{equation}\label{eq:alpha and beta}
\begin{split}
    \alpha &= \frac{\Delta \rho}{\rho^+}\left(\frac65\frac{M_t}{R_{\rm t}^3\omega^2}-\frac35\right)R_{\rm t},\\
    \beta &= R_{\rm t} -\alpha,
\end{split}
\end{equation}
where $\Delta\rho \equiv \rho^--\rho^+$.
At the surface, the Lagrangian perturbation of the pressure must vanish, i.e., $\Delta p =0$. Using Eq.~(\ref{eq:xi_r}-\ref{eq:alpha and beta}) and replacing $\rho^+$ ($\rho^-$) with $\varepsilon_{\rm t}$ ($\varepsilon_{\rm t}+\Delta \varepsilon$), we finally obtain 
\begin{equation}\label{eq:omega NT limit}
    \omega^2 = \frac{(6/5)(\Delta\varepsilon/\varepsilon_{\rm t})(1-x^5)+1+\varsigma-\sqrt{(1+\varsigma)^2-(12/5)(\Delta\varepsilon/\varepsilon_{\rm t})(\kappa-1)(1-x^5)}}{(\varepsilon_{\rm t}+\Delta\varepsilon)/\varepsilon_{\rm t}-(2/5)(\Delta\varepsilon/\varepsilon_{\rm t})(1-x^5)}\frac{M}{R^3}.
\end{equation}

\subsection{Approximated formulas of the tidal overlap integral}\label{appendix: approx tidal overlap}
\begin{figure*}[h]
    \centering
  \begin{minipage}{0.48\linewidth}
    \centering
    \includegraphics[width=2.9in]{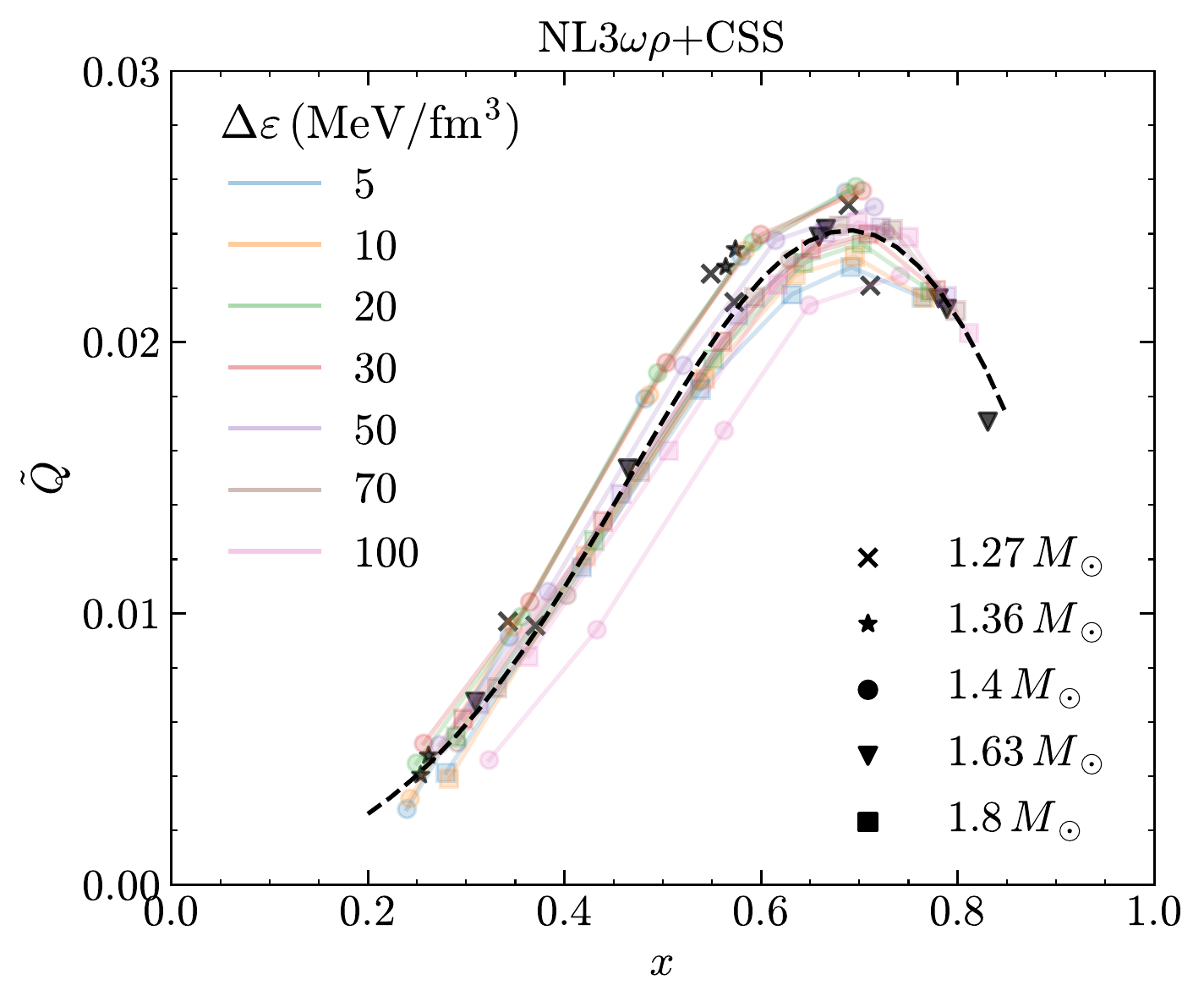}
  \end{minipage}
  \begin{minipage}{0.48\linewidth}
    \centering
    \includegraphics[width=2.9in]{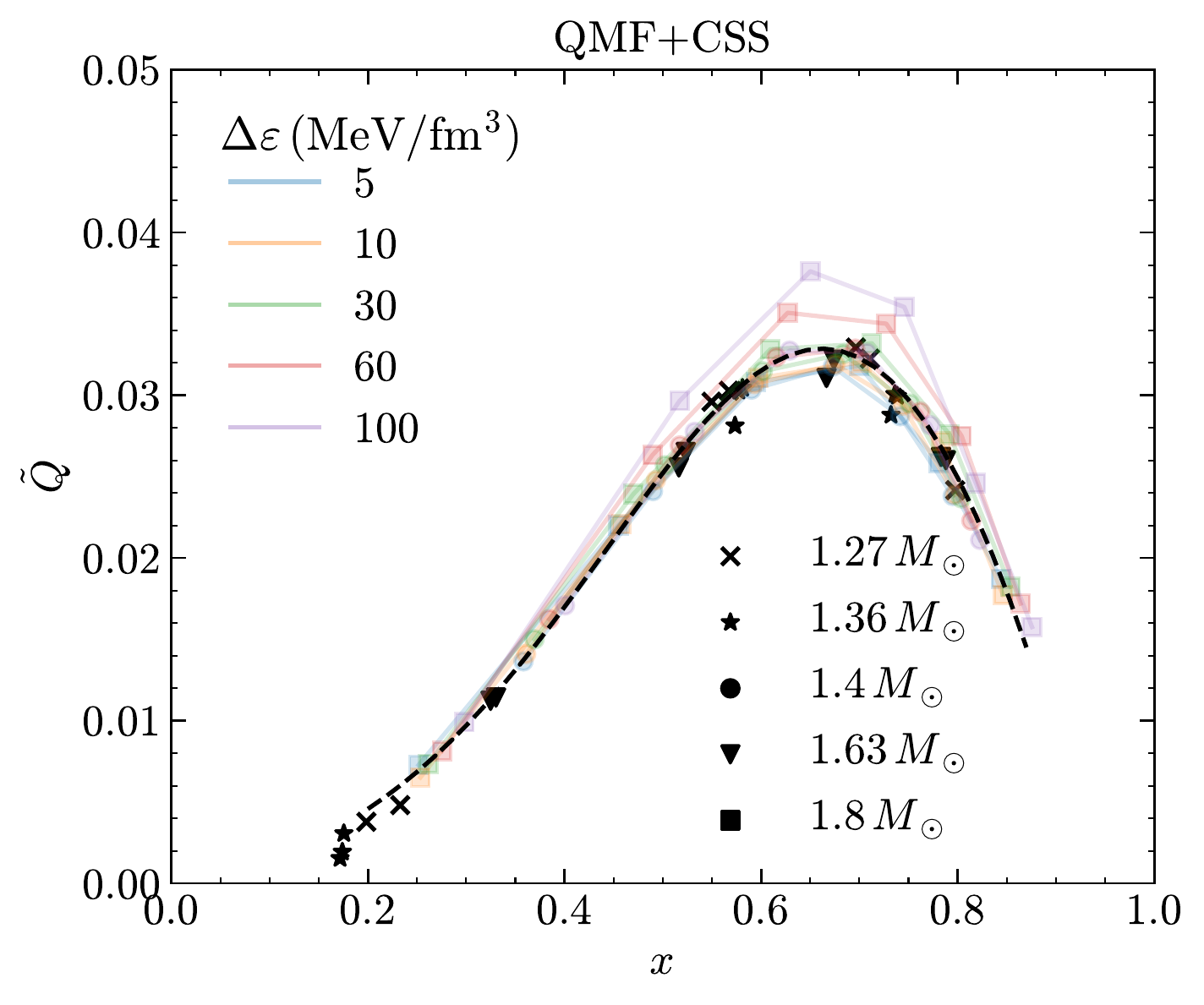}
  \end{minipage} 
    \caption{Scaled tidal overlap integral $\tilde Q$ versus $x$, together with the fitting lines for $1.4\,M_\odot$ stars. The associated fitting coefficients can be found in Table~\ref{table:coef}. 
    } 
\label{fig:Q_estimation}
\end{figure*}

We find a reliable approximation for the tidal overlap integral, as shown in Fig.~\ref{fig:Q_estimation}. We should mention that such an approximation is only valid for small transition strength, i.e., $\Delta\varepsilon\lesssim 150\,{\rm MeV/fm^3}$. However, higher transition strength ($\Delta\varepsilon\gtrsim 150\,{\rm MeV/fm^3}$) corresponds to higher mode frequencies ($\gtrsim 1200\Hz$) that we are not interested in, therefore we do not need to worry about them when using this approximation. 
By defining $\tilde Q\equiv Q[(\varepsilon_{\rm t}+\Delta\varepsilon)(\varepsilon_{\rm t}+p_{\rm t})/\varepsilon_{\rm t}^2]^{-1}M_{1.4}^{-2}$, we find
\begin{equation}\label{eq:Q approx}
    \tilde Q  \simeq \frac{cx(1-dx)}{e^{a(1-x)}+b}.
\end{equation}
We fit the data points of the $1.4\,M_\odot$ stars and the resulting fitting coefficients are presented in Table~\ref{table:coef}. It is found the estimation errors are less than $\sim25\%$.

\begin{table}
\centering
\caption{Fit parameters of the tidal overlap integral for $1.4\,M_\odot$ stars in the functional form of Eq.~(\ref{eq:Q approx}).}
\renewcommand\arraystretch{1.5}
\begin{ruledtabular}
\begin{tabular*}{\hsize}{@{}@{\extracolsep{\fill}}lcccc@{}}
Model & a & b & c& d  \\
\hline 
NL3$\omega\rho$ (stiff) +CSS & 6.304 & 17.793 & 2.827 & 1.004  \\
\hline 
QMF (soft)+CSS & 5.726 & 12.014 & 3.183 & 1.064 
\end{tabular*}
\end{ruledtabular}
    \vspace{-0.4cm}
\label{table:coef}
\end{table}

\subsection{Detectability analysis}\label{appendix: detectability}
Following \citet{1994PhRvD..49.2658C,2017MNRAS.470..350Y}, we use the Fisher information matrix to estimate the detectability of the phase shift due to resonance, assuming that the GW signal have a large signal-to-noise ratio (SNR). The Fisher information matrix for a set of parameters $\{\theta_i\}$ is defined by 
\begin{equation}
    \Gamma_{ij} = \left(\frac{\partial h}{\partial \theta_i}|\frac{\partial h}{\partial \theta_i}\right),
\end{equation}
where the inner product is defined by
\begin{equation}
  (h_1|h_2)=2\int_0^\infty \frac{\tilde h_1^*(f)\tilde h_2(f)+\tilde h_1(f)\tilde h_2^*(f)}{S_n(f)}df,
\end{equation}
with $\tilde h$ and $S_n$ are the Fourier transform of GW strain data and the spectral noise density of the detector. The SNR is given by ${\rm SNR}=(h|h)^{1/2}$. The root-mean-square (rms) measurement error of $\theta_i$ can be written as
\begin{equation}
    \Delta \theta_i =\sqrt{(\Gamma^{-1})_{ii}},
\end{equation}
where $\Gamma^{-1}$ is the inverse of the Fisher information matrix. If the rms error of phase shift $\Delta|\delta\bar\phi|$is smaller than $|\delta\bar\phi|$, then the phase shift is detectable.

We employ the IMRPhenomD\_NTidal waveform~\citep{2017PhRvD..96l1501D}, together with the addition of phase correction given by Eq.~(5), to calculate the GW signal. 
The waveform parameters are 
\begin{equation}
    \{\theta_i\} = \{\mathcal{M},q,\Lambda_1,\Lambda_2,d_L,t_c,\phi_c,|\delta\bar\phi|,\bar f\}.
\end{equation}
Here for simplicity, we omit the angular dependence and the stellar rotation. We also fix the tidal defromability as $\Lambda_{1.4}=400$ and $\Lambda_{1.8}=300$, because it is found that $\Delta|\delta\bar\phi|$ is insensitive to the choice of tidal deformability. 
In Table~\ref{table:detect precision} we list some quantitative results of rms measurement errors for a $1.4\,M_\odot-1.4\,M_\odot$ system.

\begin{table}
\centering
\caption{The rms measurement errors of $|\delta\bar\phi|$ and $\bar f$ for a $1.4\Msun-1.4\Msun$ system.}
\renewcommand\arraystretch{1.5}
\begin{ruledtabular}
\begin{tabular*}{\hsize}{@{}@{\extracolsep{\fill}}ccccccc@{}}
$\bar f\,({\rm Hz})$ & $|\delta\bar\phi|$ & $d_L\,({\rm Mpc})$  & Detector &  SNR  & $\Delta \bar f\,({\rm Hz})$ & $\Delta|\delta\bar\phi|$ \\
\hline 
\multirow{3}{*}{500} & \multirow{3}{*}{1} & \multirow{3}{*}{100} &aLIGO & 49 &140 & 1.497 \\
  &  & & ET & 502 & 18& 0.202 \\
  & & & CE & 1495 & 11& 0.127 \\
\hline
\multirow{3}{*}{750} & \multirow{3}{*}{1} & \multirow{3}{*}{100} &aLIGO & 49 &359 &2.842 \\
  &  & & ET & 502 & 57 & 0.336 \\
  & & & CE & 1495 & 36& 0.217 \\
\hline
\multirow{3}{*}{1000} & \multirow{3}{*}{1} & \multirow{3}{*}{100} &aLIGO & 49 &885 &4.060 \\
  &  & & ET & 502 & 145 & 0.664 \\
  & & & CE & 1495 & 93& 0.429 \\
\end{tabular*}
\end{ruledtabular}
    \vspace{-0.4cm}
\label{table:detect precision}
\end{table}

\subsection{Bayesian parameter estimation for GW170817}\label{appendix: bayesian analysis}
We perform a Bayesian parameter estimation to analysis the GW170817 data by incorporating the resonant parameter $|\delta\bar\phi|$ and $\bar f$. We utilize the IMRPhenomD\_NTidal waveform for the basic hypothesis $\mathcal{H}_0$, whereas add the phase correction for the hypothesis $\mathcal{H}_1$.
The waveform parameters are given by
\begin{equation}
    \{\theta_i\} = \{\mathcal{M},q,\Lambda_1,\Lambda_2,{\chi}_{1z},{\chi}_{2z},\theta_{\rm jn},t_c,\phi_c,\Psi,|\delta\bar\phi|,\bar f\}
\end{equation}
We fix the location of the source to the position determined by electromagnetic observations~\citep{2017ApJ...848L..12A,2017ApJ...848L..28L} with $\alpha(J2000)=197.45^\circ$, $\delta(J2000)=-23.38^\circ$ and $z=0.0099$.
The priors of the parameters are chosen by following those used in \citet{2019PhRvX...9a1001A}, with the exception of the priors for $|\delta\bar\phi|$ and $\bar f$. 
For the mode resonance parameters, we've chosen priors as $|\delta\bar\phi|\sim {\rm LogU}[10^{-4},10^2]$ and $\bar f\sim {\rm U}[50,1600]\Hz$, where `U' denotes the uniform distribution and `LogU' represents the logarithmic uniform distribution.

\begin{figure}[h]     \vspace{-0.4cm}
  \centering
    \centering
    \includegraphics[width=6.5in]{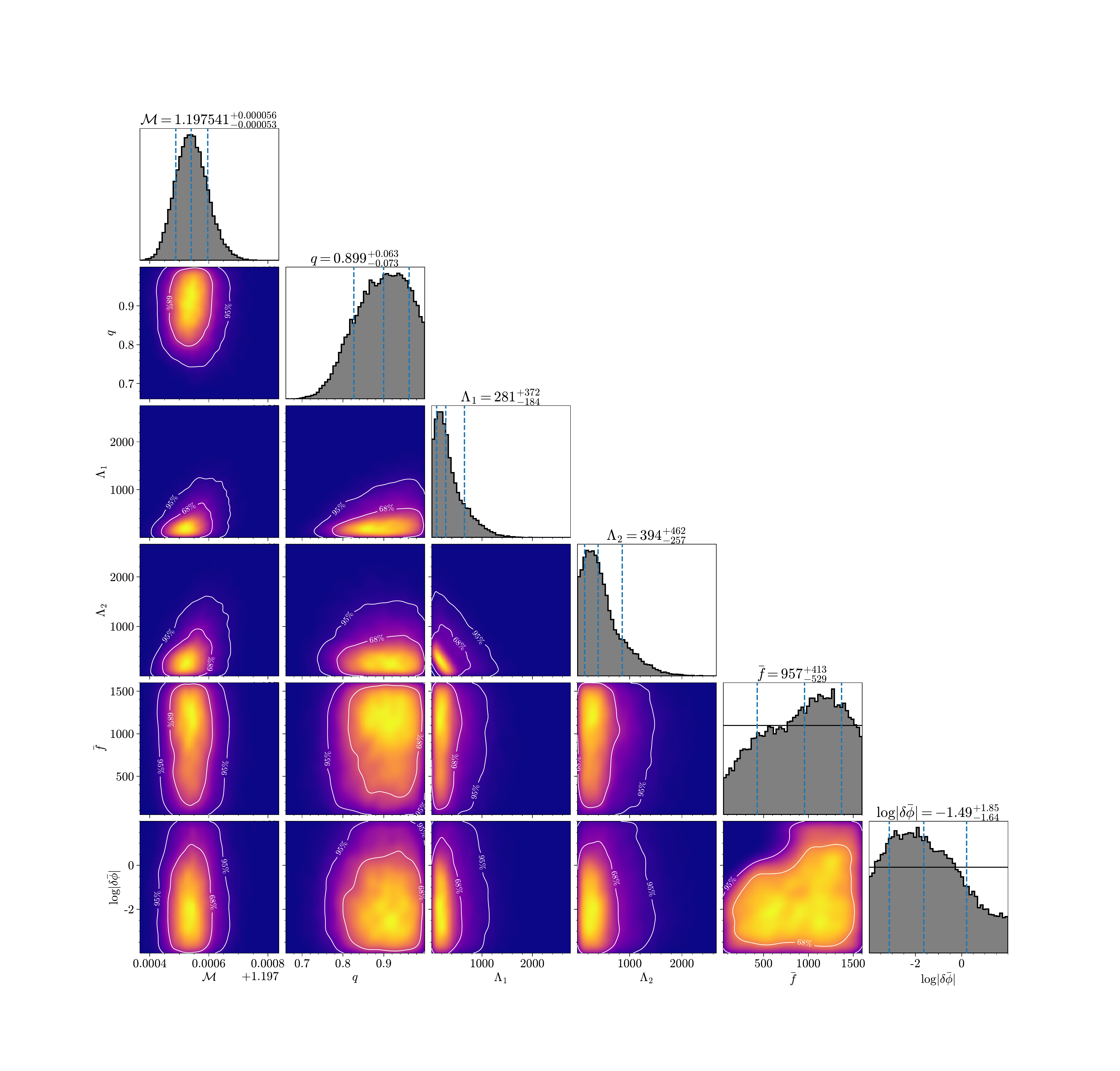}
    \caption{Posterior distributions of the waveform parameters inferred from GW170817 strain data. The horizontal lines in last two diagonal panels denote the prior distributions of corresponding parameters.
    } 
\label{fig:posterior}
\end{figure}

In Fig.~\ref{fig:posterior} we show the posterior distribution of waveform parameters. Since we use the PyMultiNest to sample from the posterior distribution, we can obtain the evidence ($\mathcal{Z}$) of each hypothesis directly. We then calculate their Bayes Factor with $\mathcal{B}_0^1=\mathcal{Z}_1/\mathcal{Z}_0$. We also calculate the Bayes Factor with the Savage-Dickey Density Ratio method~\citep{Dickey1970}, by using the posterior distribution of hypothesis $\mathcal{H}_1$ alone, as
\begin{equation}
    \frac{1}{\mathcal{B}_0^1} = \lim_{|\delta\bar\phi|\to0}\frac{p(|\delta\bar\phi||data,\mathcal{H}_1)}{p(|\delta\bar\phi||\mathcal{H}_1)}.
\end{equation}
Both methods yield similar Bayes Factors. By evaluating the evidence, we obtain $\mathcal{B}_0^1=0.72$, while using the Savage-Dickey Density Ratio method, we find $\mathcal{B}_0^1=1.11$. According to the significance quantification approach in \citet{kass1995}, when the Bayes Factor is close to unity, it indicates that the current data does not strongly favor one hypothesis over the other.

\section*{Acknowledgments}
We thank Dong Lai, Lap-Ming Lin, Yuxin Liu, Zhen Pan, Yong Gao and Zhenyu Zhu for their helpful comments.
The work is supported by the National SKA Program of China (No.~2020SKA0120300) and the National Natural Science Foundation of China (grant Nos.~12273028 and 12203017).
This research has made use of data and software obtained from the Gravitational Wave Open Science Center $https://www.gw-openscience.org$, a service of LIGO Laboratory, the LIGO Scientific Collaboration and the Virgo Collaboration. LIGO is funded by the U.S. National Science Foundation. Virgo is funded by the French Centre National de Recherche Scientifique (CNRS), the Italian Istituto Nazionale della Fisica Nucleare (INFN) and the Dutch Nikhef, with contributions by Polish and Hungarian institutes.

\bibliography{gmode.bib} 
\bibliographystyle{aasjournal}

\end{document}